\documentclass[twocolumn,prl,amsmath,amssymb]{revtex4}
\usepackage{graphicx}
\usepackage{epstopdf}
\usepackage{amsmath}
\DeclareGraphicsExtensions{.pdf,.eps,.png,.jpg,.mps}

\begin{document}

\title{Graphene-based, mid-infrared, room-temperature pyroelectric bolometers with ultrahigh temperature coefficient of resistance}
\author{U. Sassi$^1$, R. Parret$^2$, S. Nanot$^2$, M. Bruna$^3$, S. Borini$^3$, S. Milana$^1$, D. De Fazio$^1$, Z. Zhuang$^1$, E. Lidorikis$^4$, F. H. L. Koppens$^{2,5}$, A. C. Ferrari$^1$, A. Colli$^{6}$}

\affiliation{$^1$ Cambridge Graphene Centre, University of Cambridge, Cambridge CB3 0FA, UK}
\affiliation{$^2$ ICFO, Institut de Ciencies Fotoniques, The Barcelona Institute of Science and Technology, 08860 Castelldefels, Barcelona, Spain}
\affiliation{$^3$ Nokia Technologies, Broers Building, Cambridge CB3 0FA, UK}
\affiliation{$^4$ Department of Materials Science and Engineering, University of Ioannina, Ioannina 45110, Greece}
\affiliation{$^5$ ICREA, Institucio Catalana de Recerca i Estudis Avancats, Barcelona, Spain}
\affiliation{$^6$ Emberion Ltd, Cambridge CB3 0FA, UK}
\begin{abstract}
Graphene is ideally suited for photonic and optoelectronic applications, with a variety of photodetectors (PDs) in the visible, near-infrared (NIR), and THz reported to date, as well as thermal detectors in the mid-infrared (MIR). Here, we present a room temperature-MIR-PD where the pyroelectric response of a LiNbO$_3$ crystal is transduced with high gain (up to 200) into resistivity modulation for graphene, leading to a temperature coefficient of resistance up to 900$\%$/K, two orders of magnitude higher than the state of the art, for a device area of 300x300$\mu$m$^2$. This is achieved by fabricating a floating metallic structure that concentrates the charge generated by the pyroelectric substrate on the top-gate capacitor of the graphene channel. This allows us to resolve temperature variations down to 15$\mu$K at 1 Hz, paving the way for a new generation of detectors for MIR imaging and spectroscopy.
\end{abstract}
\maketitle
\section{Introduction}
Detecting thermal infrared radiation of room temperature (RT) objects (with spectral peak emittance$\sim$10$\mu$m\cite{Krus2001,Roga2011}) is increasingly important for applications in astronomy\cite{RichPT2005}, healthcare\cite{JoneIEEE1998, AhyeAHM2016}, smart energy systems\cite{Meij2008}, security\cite{RudoIEEE2008}, pollution monitoring\cite{KreuS1972}, fire sensing\cite{Evan FT1985}, automotive\cite{SimoSAA2002} and motion tracking\cite{YunS2014}. Thermal photodetectors (PDs), such as bolometers and pyroelectric detectors, have emerged as the technology of choice for RT operation, because they do not need cooling to function\cite{Roga2011,LangPT2005}. Graphene is ideally suited for photonic and optoelectronic applications\cite{BonaNP2010, FerrN2015, KoppNN2014}, with a variety of PDs in the visible\cite{XiaNN2009, MuelNP2010, EchtNC2011, EchtNL2014, EchtNL2015}, near-infrared (NIR)\cite{FerrN2015, KoppNN2014}, and THz reported to date\cite{VicaNM2012}, as well as thermal detectors in the mid-infrared (MIR)\cite{HsuNL2015, YaoNL2014, BadiNL2014}.

Pyroelectric detectors are low-cost, un-cooled thermal PDs for the MIR\cite{Krus2001,Roga2011,LangPT2005}. They are capacitor-like structures where a pyroelectric crystal is sandwiched between two metal electrodes\cite{Roga2011}. Pyroelectric crystals are materials with a T-dependent spontaneous polarization, P [C/m$^2$], i.e. surface density of bound charge\cite{LangPT2005}. Around RT, a linear relation links the T variation, $\Delta$T, with the changes of P\cite{Krus2001,Roga2011,LangPT2005}:
\begin{equation}
\label{eq1}
\Delta P = p \cdot \Delta T
\end{equation}
where p [$\mu$C/m$^2$K] is the pyroelectric coefficient (for the crystallographic direction perpendicular to the electrodes). The two metal electrodes are connected through an external load resistor $R_L$. In thermal equilibrium (dT/dt = 0), no current flows in the external circuit because P is constant and the charges on the electrodes compensate the bound charges at the pyroelectric surface\cite{LangPT2005}. However, when the detector is illuminated, the absorbed radiation heats the crystal and P changes according to Eq.\ref{eq1} \cite{LangPT2005}. The variation of the bound charge surface density will induce a current in the external circuit\cite{LangPT2005}:
\begin{equation}
\label{eq2}
I_p = A \cdot \frac{dP}{dt} = A \cdot p \cdot \frac{dT}{dt}
\end{equation}
where A [m$^2$] is the electrode area\cite{Krus2001,WhatRPP1986}. $I_p$ flows only as long as T changes (i.e. when the impinging optical power changes).

Bolometers are another class of un-cooled thermal PDs, where T variations due to incoming photons produce a change in the resistance (R) of a sensing element. This can be a thin metal layer\cite{PurkIEEE2013}, a semiconductor\cite{JeroOE1993} or a superconductor\cite{Krus2001}. Common metallic bolometers for RT operation are made of Ti\cite{JuSPIE1999}, Ni\cite{BrocJOPSA1946} or Pt\cite{PurkIEEE2013}. Polysilicon\cite{Krus2001, Roga2011}, amorphous silicon\cite{TissoSPIE1994} or vanadium oxide\cite{JeroOE1993} are usually exploited for semiconducting bolometers. For fixed bias $V_d$, the resistance change of the sensing element translates in a measurable change in current (I). The TCR [$\%$/K] is a key performance indicator for a bolometer, and is defined as\cite{Roga2011}:
\begin{equation}
\label{eq3}
TCR(R_0) = \frac{1}{R_0} \cdot \frac{dR}{dT} = \frac{1}{I_0} \cdot \frac{dI}{dT}
\end{equation}

The TCR represents the percentage change in resistance per Kelvin around the operating point $R_0$, and corresponds in module to the normalized current change per Kelvin around the operating current $I_0$ (Eq. \ref{eq3}). The TCR in metallic bolometers is$\sim$0.4$\%$/K\cite{Roga2011}, whereas for semiconducting bolometers it is higher$\sim$2-4 $\%$/K\cite{Krus2001,Roga2011}. It follows that the output of a bolometer (measured current) is proportional to T, in contrast to the output of a pyroelectric detector (measured current) that depends on the derivative of T, see Eq.\ref{eq2}\cite{WhatRPP1986}. However, while the TCR of a bolometer does not depend on device area, in pyroelectric detectors the output current is a function of the electrode size, as for Eq.\ref{eq2}\cite{WhatRPP1986}. Larger electrodes allow the collection of more charge, increasing the pyroelectric current, therefore leading to a larger signal.

These differences have an impact on the suitability of both technologies for different applications. Because pyroelectric detectors are alternating current (ac) devices that rely on a variable impinging radiation, they require a chopper at 25-60Hz\cite{LangPT2005} to detect stationary objects, and are thus preferentially used to detect moving targets (e.g., for automatic lighting systems\cite{Meij2008}, electrical outlet turn-off\cite{YunS2014}, unusual behavior detection\cite{HaraPNN2002}, home invasion prevention\cite{LangPT2005}, etc.), where they are not only able to detect the presence of a warm bodies\cite{Roga2011}, but also to extrapolate parameters like distance, direction, or speed of movement\cite{YunS2014}. Such information can be obtained by processing the analog signals of only a couple of large ($\sim$1cm$^2$) detectors\cite{YunS2014}. On the other hand, bolometers can be scaled to smaller sizes without any loss in TCR to make arrays of pixels for stationary imaging. Resistive micro-bolometers used in high-resolution thermal cameras range from 17x17 to 28x28 $\mu$m$^2$ in size\cite{Roga2011}.

Refs.\cite{VoraAPL2012, YanNN2012, HanSR2013} reported graphene-based bolometers working at T$<$10K. However, these devices are not viable for practical applications in the mass market (as for examples above), where RT operation is needed. At RT, single-layer graphene (SLG) is not competitive as the sensing element of a bolometer, as it shows a maximum TCR$\sim$0.147$\%$/K\cite{LangPT2005}, lower than both the metallic and semiconducting bolometers discussed above. Ref.\cite{BaeN2015} used reduced graphene oxide films with TCR$\sim$2.4-4$\%$/K at RT, whereas Ref.\cite{Hazr2013} exploited vertically aligned graphene nanosheets to produce IR bolometers with TCR$\sim$11$\%$/K at RT, the largest, to date, for carbon nanomaterials\cite{Hazr2013}. In these films, however, conduction is modulated by thermally assisted electron hopping between different sheets or localized defect sites\cite{BaeN2015} and not by any SLG intrinsic property.

SLG, however, can play a key role when integrated in pyroelectric devices. SLG can be used as a transducer for the pyroelectric polarization due to its field-effect response\cite{HsieAPL2012}. If a graphene field-effect transistor (GFET) is fabricated on a pyroelectric substrate, the channel resistance is modulated by the substrate polarization and can thus represent a direct T readout. This is the electrical equivalent of a bolometric response with area-independent TCR (the charge density of the pyroelectric 'gate' is constant, and does not depend on the size and shape of the graphene channel). Ref.\cite{HsieAPL2012} reported TCR$\sim$6$\%$/K for GFETs on lead zirconate titanate (PZT), a material with one of the highest pyroelectric coefficient known to date (up to 780$\mu$C/m$^2$K)\cite{GuoAPL2007}. This indicates that the pyroelectric charge density generated by PZT underneath the SLG channel is still too limited to significantly outperform state-of-the-art bolometers\cite{HsieAPL2012}.

Here, we demonstrate a RT pyroelectric bolometer for MIR with ultrahigh TCR. Our PD is a two-terminals device whose resistance changes proportionally to T, like a bolometer. Internally, the PD comprises a floating metallic structure that concentrates the charge generated by the pyroelectric substrate over an integrated GFET. Since charge cannot escape from the floating structure (i.e. there is no load resistor), the PD can be operated in direct current (dc) and there is no need for chopping. The total pyroelectric charge generated upon a variation in T increases with area, delivering TCRs up to 900$\%$/K for a footprint of 300x300$\mu$m$^2$, 2 orders of magnitude larger than state-of-the-art IR detectors having any similar or larger area\cite{Krus2001, Roga2011,BaeN2015, Hazr2013}. The TCR scaling is sub-linear for smaller footprints. We discuss the origin of this behavior and conclude that our device performance is competitive even in the limit of very small pixels.
\section{Results and Discussion}
\begin{figure*}
\includegraphics[width=170mm]{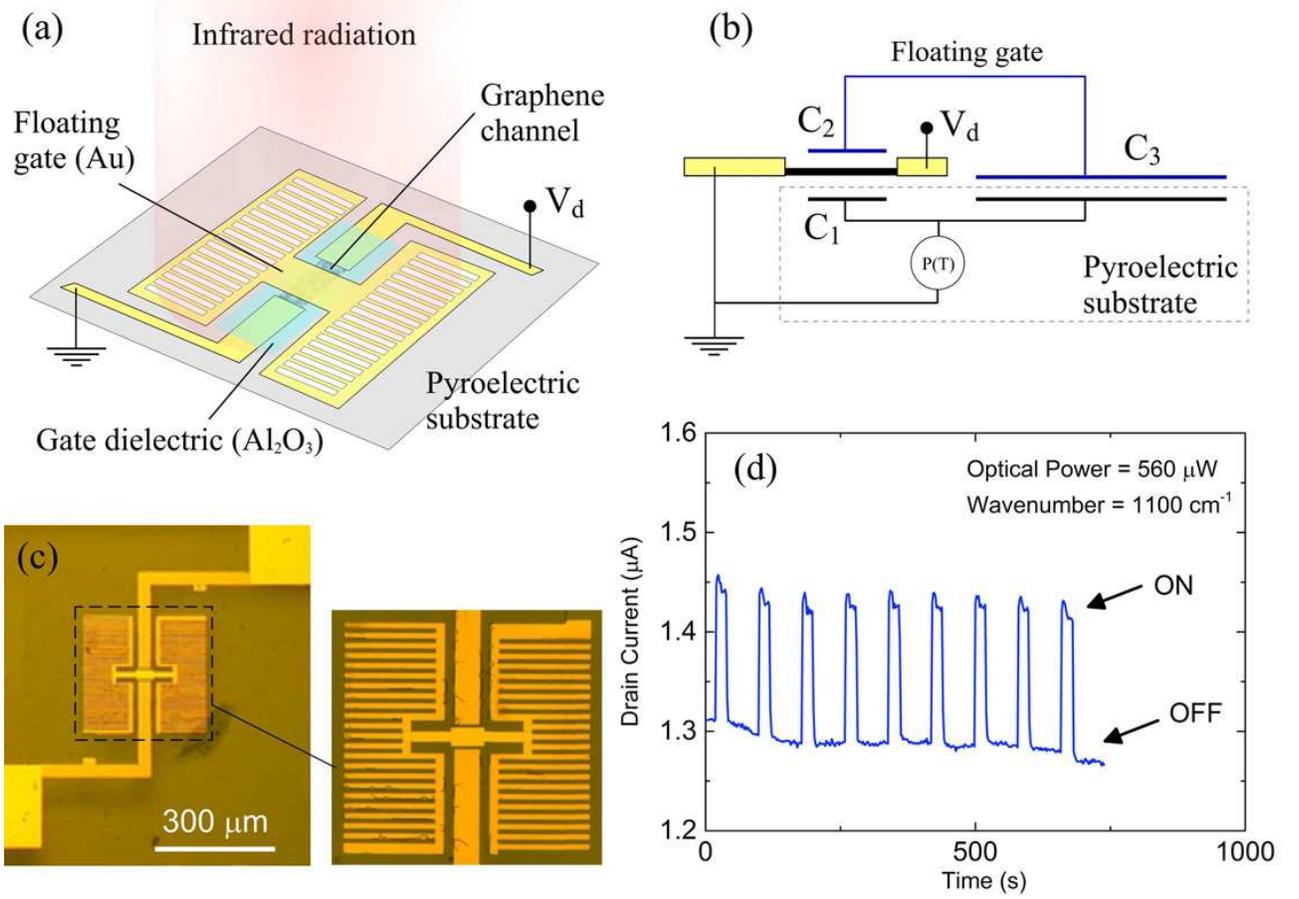}
\caption{(a) Schematic of an individual graphene pyroelectric bolometer, where the conductance of a SLG channel is modulated by the pyroelectric substrate and by a floating gate. This is driven by two metallic pads in contact with the substrate, with a total area much larger than the overlap with the SLG channel. Such pads can be either uniform or patterned. (b) Circuit diagram for the device in (a). (c) Optical image of a device with lateral pads patterned as electrically-connected finger-like structures. (d) Response at 1100 cm$^{-1}$ ($\sim$9$\mu$m) over several ON/OFF cycles induced by a manual shutter. The laser spot size is 300 $\mu$m. The drain current is measured for a 10mV drain voltage ($V_d$).}
\label{fig:Fig1}
\end{figure*}
Figs.\ref{fig:Fig1}a,b show the layout of a single device and the corresponding electrical model. A SLG channel with source and drain contacts is fabricated on a pyroelectric substrate: 500$\mu$m-thick z-cut LiNbO$_3$ (LN) crystal, as described in Methods. A 10 nm-thick Al$_2$O$_3$ dielectric layer isolates the SLG from an H-shaped floating Au structure designed to overlap the oxide-coated SLG in the center, while lateral pads are placed in direct contact with the substrate. Both uniform and patterned Au pads have been studied, the latter in the form of finger-like structures (in order to enhance light absorption at selected wavelengths, see Methods). The design is such that the SLG channel conductivity can be modulated by a dual-gate capacitive structure. From the bottom, there is the pyroelectric polarization (and associated electric field) generated directly by the substrate ($C_1$ in Fig.\ref{fig:Fig1}b), which we refer to as the 'direct effect' on graphene conductivity (previously exploited in Ref.\cite{HsieAPL2012}). From the top, there is a gate $C_2$ connected in series with capacitor $C_3$ as a floating circuit branch, with $C_3>C_2$. The perimeter of the pads defining $C_3$ sets the overall pixel size, from which only the source and drain contacts stem out to interface with the measurement electronics.

In first approximation, the generated pyroelectric charge $\Delta$Q is uniformly distributed on the substrate upon a T variation\cite{Krus2001, Roga2011}. Therefore, the direct effect from $C_1$ does not depend on the channel area $A_{C1}$, since the bottom gate field depends on the pyroelectric polarization, which is constant over any area. For the floating gate in Fig.\ref{fig:Fig1}b, $\Delta$Q accumulating on $C_3$ depends on area as (from Eq.\ref{eq1}) $\Delta Q=p \Delta T A_{C3}$. Being the structure electrically floating, the same $\Delta$Q will be stored on $C_2$, generating an effective top-gate voltage (in module):
\begin{equation}
\label{eq4}
\Delta V_{TG}=\frac{\Delta Q}{C_2}=\frac{p \Delta T t}{\epsilon_0 \epsilon_r} \cdot \frac{A_{C3}}{A_{C2}}
\end{equation}

where $C_2=\epsilon_0 \epsilon_r A_{C2} /t$, $\epsilon_0$ and $\epsilon_r$ are the vacuum and relative permittivity, and t is the oxide thickness. For fixed t and $\Delta$T, $A_{C3}/A_{C2}$ controls the amplification through the top gate voltage driving the GFET, modulating the graphene drain current.

Fig.\ref{fig:Fig1}c shows an optical micrograph of a device with patterned pads. We illuminate this device with MIR radiation at 1100cm$^{-1}$ ($\sim$9$\mu$m) using a laser spot matching the pixel size (300x300$\mu$m$^2$). The resulting modulation of the channel drain current is shown in Fig.\ref{fig:Fig1}d over 9 ON/OFF laser cycles. A responsivity$\sim$0.27mA/W is obtained, for a drain current in the dark ($I_{OFF}$)$\sim$1.3$\mu$A ($V_d$=10mV). The responsivity can be increased by applying a larger $V_d$, which in turns increases $I_{OFF}$. The normalized current responsivity in [$\%$/W] is defined as:
\begin{equation}
\label{eq5}
R_{ph,N} = \frac{\frac{I_{ON}-I_{OFF}}{I_{OFF}}\cdot 100}{P_{in}}
\end{equation}
where $I_{ON}$ is the current under illumination and $P_{in}$ is the optical power of the incoming radiation. This is a better parameter to compare photoconductive detectors. $R_{ph,N}$ in Fig.\ref{fig:Fig1}d is$\sim2\cdot10^4\%$/W, over two orders of magnitude higher than in Ref.\cite{HsieAPL2012}, where only the direct effect was exploited ($\sim1.2\cdot10^2\%$/W)\cite{HsieAPL2012}.

\begin{figure*}
\includegraphics[width=170mm]{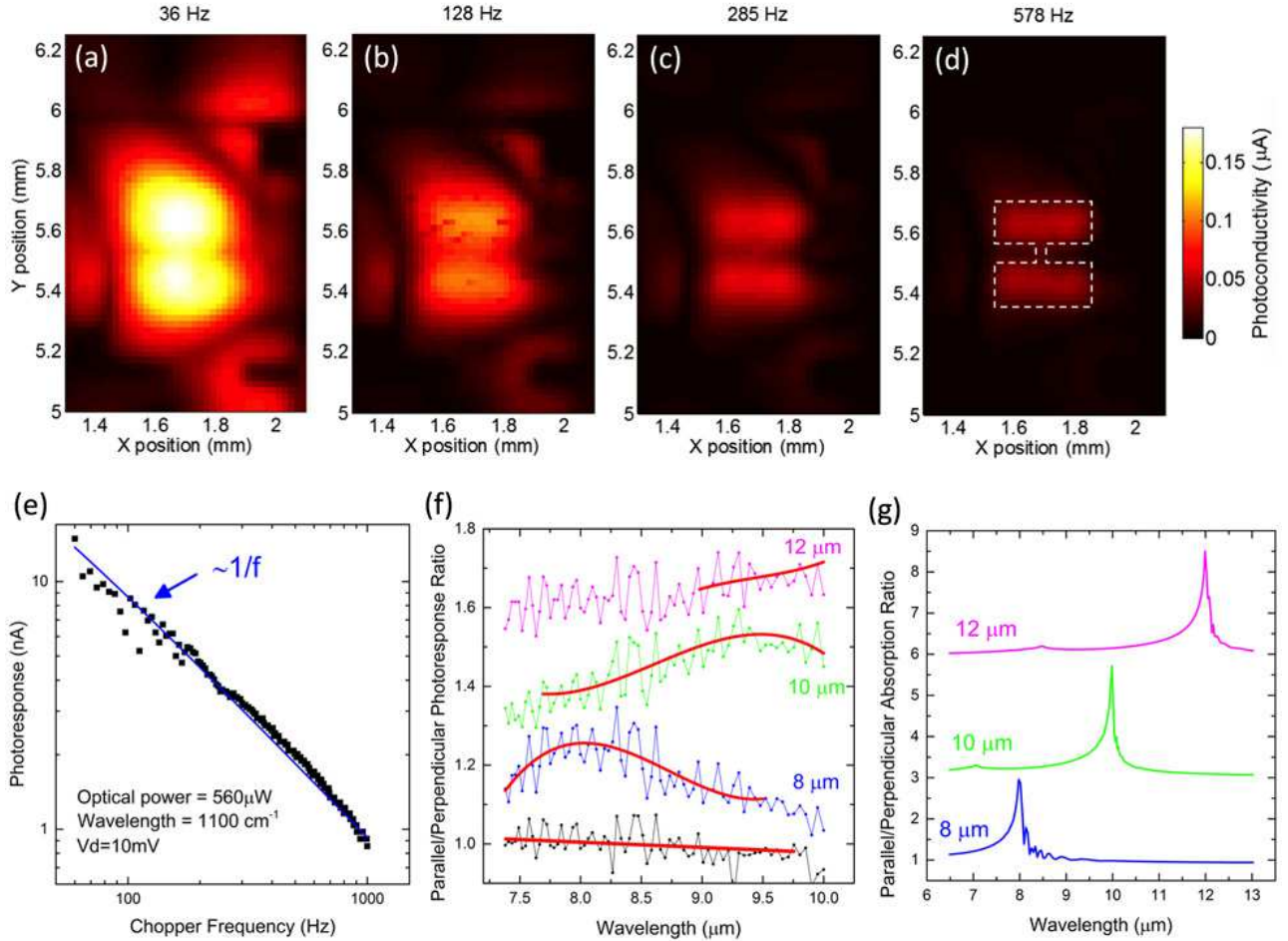}
\caption{(a)-(d) Photocurrent profiles of a representative device for different f and a beam of 1.8mW at 1100cm$^{-1}$. The peak intensity decreases at higher f, but the map is more resolved. The dashed white lines in (d) indicate the location of all device features, rotated 90$^\circ$ with respect to Fig.\ref{fig:Fig1}c. (e) Dependence of the locked-in photoresponse with f for a fully-illuminated device (laser diameter=300$\mu$m), showing a 1/f scaling above 60 Hz. (f) Wavelength dependence of the photoresponse for a device patterned as in Fig.\ref{fig:Fig1}c. The pitch of the fingers varies between 8 and 12$\mu$m. Peaks are observed in the photoresponse ratio between parallel and perpendicular polarized light and their position changes with pitch (the red lines are guides to the eye). The flat photoresponse ratio for a device with uniform Au pads (black data points) is also shown as reference. (g) Simulated total absorption for the devices measured in (f). Predicted peaks in the parallel/perpendicular absorption ratio match those measured for the photoresponse.}
\label{fig:Fig2}
\end{figure*}

By reducing the laser spot to $\sim$10$\mu$m and using a lock-in, we produce photocurrent maps of a single pixel to assess where the maximum  signal is generated (Figs.\ref{fig:Fig2}a-d).
At the slowest chopper frequency (f=36Hz, Fig.\ref{fig:Fig2}a) the photocurrent map shows two broad peaks ($\sim$700$\mu$m), which largely overlap and extend beyond the pixel area. When f is increased, the two peaks become progressively resolved until, above 500 Hz (Fig.\ref{fig:Fig2}d), they match the location of the lateral pads defining the pixel. These are not necessarily the areas where the strongest absorption occurs, but those where T changes are detected providing the highest photocurrent $I_{ph}=I_{ON}-I_{OFF}$. Figs.\ref{fig:Fig2}a-d also show that the signal decreases at higher f. To better quantify this trend, we plot $I_{ph}$ upon full illumination (300x300$\mu$m$^2$ spot size, Fig.\ref{fig:Fig2}e) as a function of f. $I_{ph}$ scales linearly with 1/f and is measurable up to 1kHz. This PD can be described by a thermal model, see Methods. This confirms the most intuitive interpretation of the measurements. When an increase in illumination time per cycle (i.e. a reduction in f) results into a proportional T increase, the system is far from a dynamic equilibrium with the thermal sink (the chip carrier) within a single cycle. This behavior is observed for f$>$60Hz. Also, since at slow chopping speeds (Fig.\ref{fig:Fig2}a,b) there is more time to approach dynamic equilibrium, including to laterally spread away from the illuminated spot, the T within the pixel becomes more homogeneous, resulting in blurred photomaps.

All results presented so far are for monochromatic illumination at 1100cm$^{-1}$ ($\sim$9.1$\mu$m,136meV). Fig.\ref{fig:Fig2}f plots the wavelength dependence of the photocurrent for devices with lateral pads patterned with a finger-like design (as in Fig.\ref{fig:Fig1}c) with different pitches. The fill ratio (i.e. Au finger area/total available pad area) is kept constant at 0.5, meaning that, e.g., fingers with an 8$\mu$m pitch are 4$\mu$m-wide and separated by a 4$\mu$m gap. While for a uniform Au pad the photoresponse to parallel and perpendicular light is the same for all wavelengths (black data in Fig.\ref{fig:Fig2}f), for patterned pads a peak arises in the parallel/perpendicular photoresponse ratio at the wavelength that matches the fingers pitch.

We then simulate the total parallel/perpendicular absorption for finger-like Au structures on thick LN (Fig.\ref{fig:Fig2}g, see Methods for details). This shows that a peak is expected at the wavelength corresponding to the fingers pitch. This is consistent with Fig.\ref{fig:Fig2}f, as more absorption results into a larger T increase, therefore a larger signal. In the calculations we assume that all light entering the bulk LN substrate is eventually absorbed, contributing to the T rise. Parallel-polarized light gets more absorbed overall (i.e. in the Au fingers plus LN substrate) than the perpendicular-polarized light, despite the fact that at the resonant frequency the perpendicular-polarized light is absorbed more inside the Au fingers (see Methods). The reason for this is that resonant absorption in the fingers is also accompanied by resonant reflection (RFL), which lowers the overall delivery of light into the Au/LN system as 1-RFL. Things would change if these devices were fabricated on pyroelectric layers$\sim$1$\mu$m thick (rather than a 500$\mu$m LN crystal). The absorption of the resonant structures would become dominant over the intrinsic absorption of the substrate and reflectance would play a minor role. Fig.\ref{fig:Fig2}f proves that our device layout is well suited for the implementation of photonic structures to engineer photon absorption and that a spectrally-selective MIR response is feasible.

The data in Figs.\ref{fig:Fig2} suggest that even better results could be obtained for devices with an optimized thermal management compared to that offered by a bulk 500$\mu$m-thick pyroelectric substrate. Isolated pixels with lower heat capacity, using thin suspended bridges or membranes (as routinely done for microbolometers, with typical thickness$\sim$1$\mu$m\cite{Roga2011}), would further boost responsivity, speed and wavelength selectivity.

We now consider the performance as local thermometer, independent from the conversion of photons to T (linked to the emissivity and to the thermal properties of the device, such as thermal capacity and thermal conductance\cite{Roga2011,WhatRPP1986}). Each pixel is a two terminal device whose resistance represents a readout of the local T. As for bolometers\cite{Roga2011}, we consider the TCR as the figure of merit. Being a normalized parameter, the TCR does not depend on $V_d$. For our devices it depends on pixel area, so we will link each TCR value to the size of the corresponding pixel. Furthermore, since our variations in resistance are the result of a gain mechanism, we consider how they compare to the device noise. We introduce the noise equivalent substrate temperature (NEST), i.e. the pixel T change needed to produce a signal equal to the amplitude of the noise. This is not to be confused with the noise equivalent T difference (NETD)\cite{Krus2001}, often used to indicate the smallest detectable T change in an IR-emitting body imaged by a PD, also dependent on the photon-to-T conversion\cite{Krus2001}.
\begin{figure*}
\includegraphics[width=170mm]{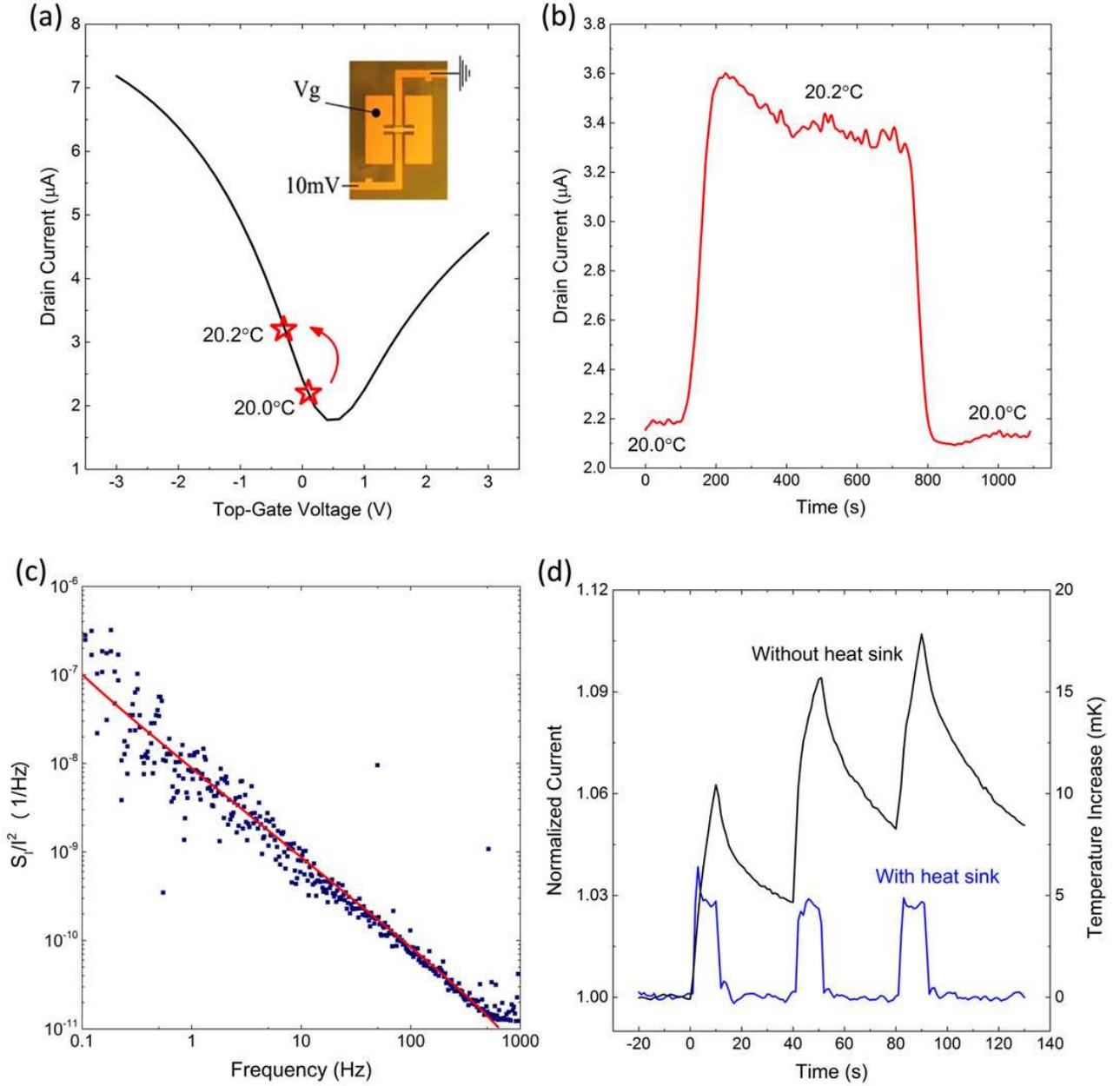}
\caption{(a) Transfer characteristics of a typical device acquired by driving $V_g$. (b) T response of the same device under floating gate conditions. Once plotted over the GFET Dirac curve in (a), the change in drain current shows that a T change of 0.2$^\circ$C produces a $V_g$ = -0.44V. (c) Normalized noise spectrum density for a representative device at constant T, showing the typical 1/f behavior observed for SLG channels. (d) Normalized current response of a SLG pyroelectric bolometer to thermal body radiation (a human hand at a distance$\sim$15cm). The local T increase is estimated from the device TCR. The T transients change in amplitude and speed according to the heat sink efficiency.}
\label{fig:Fig3}
\end{figure*}

In Fig.\ref{fig:Fig3}a,b we investigate the thermo-electrical characteristics of a representative device (pixel size: 100x100$\mu$m$^2$, $A_{C3} /A_{C2}$=22) by placing the sample on a chuck with T control. Since for these measurements the sample is in thermal equilibrium in the dark and photon absorption plays no role, uniform Au pads are chosen to accommodate (when needed) an electrical probe on the gate pads and apply an external $V_g$. Fig.\ref{fig:Fig3}a plots a typical Dirac curve for a GFET at T=20$^\circ$C. Note that, when an active electrical probe is connected to the gate pad, this neutralizes any charge coming from $C_3$ and the transfer characteristics are equivalent for all devices ($A_{C2}$ is fixed at 22x20$\mu$m$^2$). The vast majority of our devices are slightly p-doped (hole density$\sim$2.5-3$\cdot$10$^{12}$cm$^{-2}$, in agreement with our Raman data, see Methods), which is ideal to achieve a maximum signal upon heating. This is linked to our choice of placing the SLG on the positive face of z-cut LN, where heating reduces the net dipole moment, equivalent to a negative $V_g$\cite{LangPT2005}.

After removing the gate probe to leave the gate structure floating, we monitor the GFET drain current while T is raised by 0.2$^\circ$C, kept constant for 10 minutes, then decreased to its original value. Fig.\ref{fig:Fig3}b shows that the drain current increases by$\sim$50$\%$ for a 0.2$^\circ$C T change (TCR$\sim$250$\%$/K), it is stable over time, and then returns to its original value with negligible hysteresis. The red star markers in Fig.\ref{fig:Fig3}a show how the SLG conductivity evolves when the gate is thermally driven as in Fig. \ref{fig:Fig3}b. This stable DC response over several minutes indicates that no appreciable leakage occurs through the pyroelectric crystal and/or the GFET gate within a practical measurement timeframe. The initial bump to 3.6$\mu$A in Fig.\ref{fig:Fig3}b is due to the small overshoot of the chuck T at the end of the ramp.

In order to evaluate the NEST we measure the normalized noise power spectrum for a representative device (Fig.\ref{fig:Fig3}c). The spectrum is dominated by 1/f noise up to 1kHz and closely resembles those previously measured for SLG devices\cite{BalaNN2013}. The channel area (LxW) normalized noise ($S_I/I^2$)(LxW) is$\sim$5$\cdot$10$^{-7}$ $\mu$m$^2$/Hz at 10Hz (considering our 20x30$\mu$m$^2$ channel), which slightly exceeds the typical range ($\sim$10$^{-8}$-10$^{-7}\mu$m$^2$/Hz) reported for SLG devices on SiO$_2$\cite{BalaNN2013}. Considering a pixel size of 100x100 $\mu$m$^2$ and a TCR$\sim$214$\%$/K, we get for the device in Fig.\ref{fig:Fig3}b a NEST$\sim$40$\mu$K/Hz$^{1/2}$ at 1Hz (NEST=$(S_I/I^2)^{1/2}$/TCR). For the biggest pixel size (300x300$\mu$m$^2$, TCR$\sim$600$\%$/K), the minimum NEST is$\sim$15$\mu$K/Hz$^{1/2}$. For the MIR PD in Fig.\ref{fig:Fig1}, this T resolution translates into a noise equivalent power (NEP)$\sim$5$\cdot$10$^{-7}$W/Hz$^{1/2}$ at 1Hz (NEP=$(S_I/I^2)^{1/2}$/R$_{phN}$), almost one order of magnitude better than Ref.\cite{KulkOM2015} and a very promising number considering the limitations in terms of thermal conductivity and mass.

To appreciate what these numbers mean in practice, we plot in Fig.\ref{fig:Fig3}d the photoresponse of a large device (300x300$\mu$m$^2$, TCR$\sim$600$\%$/K) illuminated by the IR radiation emitted by a human hand at a distance$\sim$15cm. In one test the sample sits on a large (200mm diameter) metal chuck (with heat sink, blue data), and in another it is placed in a concave plastic box that keeps it suspended, thus more thermally isolated (without heat sink, black data). Without heat sink, the PD heats up more (hence larger device responsivity), but its response and recovery are much slower. Even with heat sink, the proximity of the hand is easily detected. The saturation signal is$\sim$3$\%$, corresponding to a T increase$\sim$5mK. For the sake of comparison, to the best of our knowledge, the only RT SLG detector working at 10$\mu$m and able to allow human hand detection is described in Ref.\cite{HsuNL2015}. This was fabricated on a suspended and very thin ($<$1$\mu$m) SiN membrane, measured in vacuum and with a lock-in with 10s integration time\cite{HsuNL2015}. Here we achieve the same result on a bulk (500$\mu$m-thick) substrate with a resistive measurement in air for 200ms, indicating that our SLG-based pyroelectric bolometer can provide far better performance (in terms of responsivity and speed) in equivalent conditions.
\begin{figure}
\includegraphics[width=83mm]{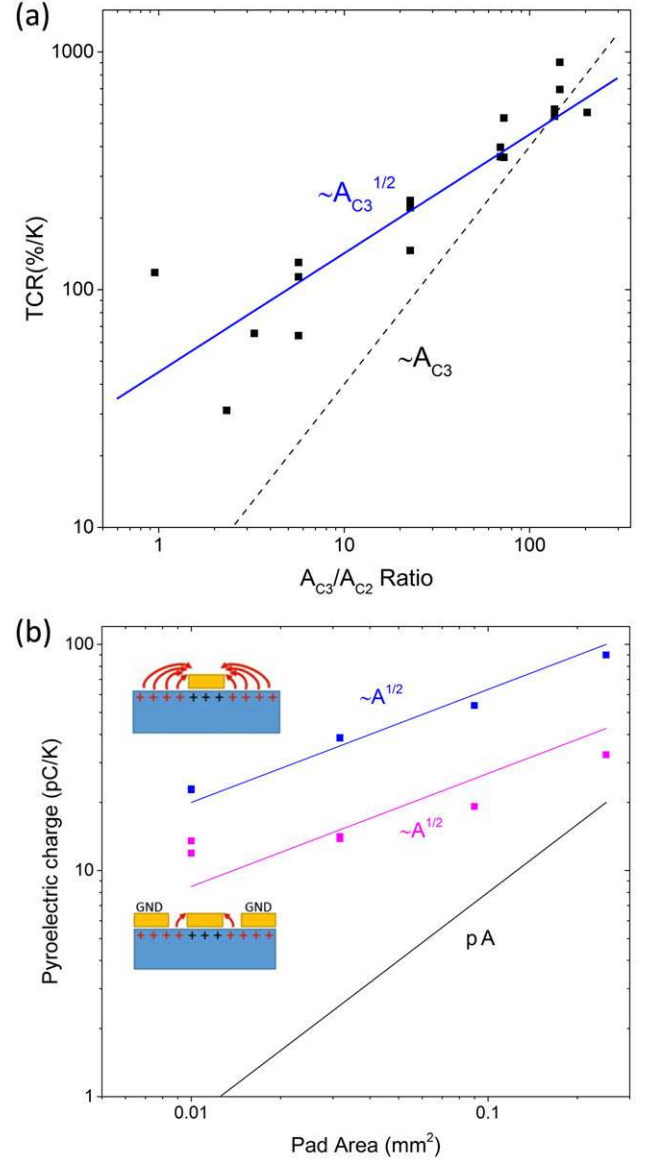}
\caption{(a) TCR for 18 devices with different $A_{C3}/A_{C2}$ ($A_{C3}$ varies with $A_{C2}$ constant), extracted from thermo-electrical measurements as in Fig.\ref{fig:Fig3}b. For decreasing $A_{C3}$, the TCR follows a square root law (blue line), instead of the linear dependence predicted by the model\cite{Roga2011} (black dashed line). (b) Integrated pyroelectric charge per Kelvin for unscreened (blue) and screened (magenta) Au pads on z-cut LN. A square root scaling law with area is found in both cases, which explains the behavior observed in (a). Even screened, small pads still offer a significant enhancement of the pyroelectric charge compared to what expected from the model\cite{Roga2011} ($\sim$pA, black line).}
\label{fig:Fig4}
\end{figure}

Finally, we discuss how our TCR scales with $A_{C3}/A_{C2}$. Fig.\ref{fig:Fig4}a plots the measured TCR for 18 devices fabricated by keeping $A_{C2}$ constant (22x20$\mu$m$^2$) and varying $A_{C3}$ from 25x25$\mu$m$^2$ to 300x300$\mu$m$^2$. Under our design assumptions (and disregarding $C_1$), the TCR should be proportional to the pyroelectric charge generated by $C_3$, hence, from Eq.\ref{eq1} one would expect a linear relation TCR$\propto A_{C3}$. Our data, however fit a square root dependence, indicated by the blue line. This behavior cannot be explained by invoking the direct effect, whose contribution appears on a much smaller scale (TCR$\sim$5$\%$/K, see Methods). Rather, we have to consider that the pyroelectric substrate does not end at the pads edge. Such pads are thus not driven just by the crystal below them (as assumed by a linear dependence on area), but can also be affected by the exposed polarization of their surrounding areas (an effect scaling linearly with perimeter, hence the square root dependence on area). To better quantify this behavior, we prepare Au pads of different size (0.01-0.3mm$^2$) on our pyroelectric substrates and measure the total charge generated upon heating (Fig.\ref{fig:Fig4}b). This is accomplished by placing an electric probe on each pad, connecting the probe to ground and integrating the pyroelectric current flowing through the probe over the whole temperature ramp. In one case, the pads are kept isolated on the 1x1cm$^2$ pyroelectric surface and independent from each other. In another case, the whole surface around the pads is coated with Au and grounded during all measurements (with only a gap of 5$\mu$m uncoated around the pads). This is meant to suppress any contribution from areas beyond the pad footprint. Fig.\ref{fig:Fig4}b, however, shows that a square root dependence on pad area is observed in both cases. The total pyroelectric charge decreases by a factor$\sim$2-3 upon screening, but is still above what would be expected from the model\cite{Roga2011}, Q/$\Delta$T=p A, even for relatively large pads (using p=77$\mu$C/m$^2$K, as measured for a LN sample fully covered with Au and consistently with literature\cite{Roga2011,LangPT2005,Wong2002}). This result has major technological implications, since it proves that a substantial contribution to the observed TCR enhancement for small $A_{C3}$ arises within the first few microns from the pad edge. It is then possible to harvest an enhanced pyroelectric charge in a dense array of small pixels, with only a gap of few microns separating two adjacent devices.

In principle, $A_{C3}/A_{C2}>10$ is desirable to deliver TCRs up to 900$\%$/K (Fig.\ref{fig:Fig4}a, see also Methods), but this upscaling is bound by the maximum gate voltage variation allowed for the GFET (dynamic range). When probed electrically (Fig.\ref{fig:Fig3}a), our GFETs show no gate leakage up to $\pm$5V ($\sim$5MV/cm). Beyond this value dielectric breakdown can occur. This determines the maximum thermal shock a device can sustain without failing, inversely proportional to the TCR. However, it is not a concern if the environment T is drifting on a time scale much larger than the measurement timeframe, e.g. during a day/night indoor T cycle. While internal pyroelectric leakage can be neglected over a few minutes, it can still discharge a device over longer times. This will always leave the GFET at the best operating point to respond to sudden signals (see Methods). If one wants to scale down the pixel size while maintaining the same area ratio, the channel area must be decreased accordingly. Since the 1/f noise scales with channel area\cite{BalaNN2013}, we expect the GFET noise to increase and cancel the benefit of a large TCR when the NEST is evaluated. Fig.\ref{fig:Fig4} however shows that the TCR scales sub-linearly with area. For pixels approaching the scale required for high resolution (20x20 $\mu$m$^2$) IR cameras\cite{Roga2011}, it is better to make a large (several $\mu$m) channel and accept $A_{C3}/A_{C2}<$10 because the small price paid in terms of TCR will be more than compensated by lower noise, less critical lithography, and a detector more resilient to sudden thermal shocks.
\section{Conclusions}
We presented a graphene-based pyroelectric bolometer operating at room-temperature with TCR up to$\sim$900$\%$/K for a device area $\sim$300x300$\mu$m$^2$ able to resolve temperature variations down to 15$\mu$K at 1Hz. For smaller devices the TCR scales sub-linearly with area, due to an enhancement of the collected pyroelectric charge in close proximity to the metallic edges. When used as mid infrared detectors, our devices deliver very promising performance (in terms of responsivity, speed and NEP), even on bulk substrates, and are capable to detect warm bodies in their proximity. Spectral selectivity can be achieved by patterning resonant structures as part of the pixel layout. This is competitive on a number of levels, ranging from high-resolution thermal imaging (small pixels) to highly-sensitive spectroscopy in the mid- and far-IR (large pixels).
\begin{figure}
\includegraphics[width=90mm]{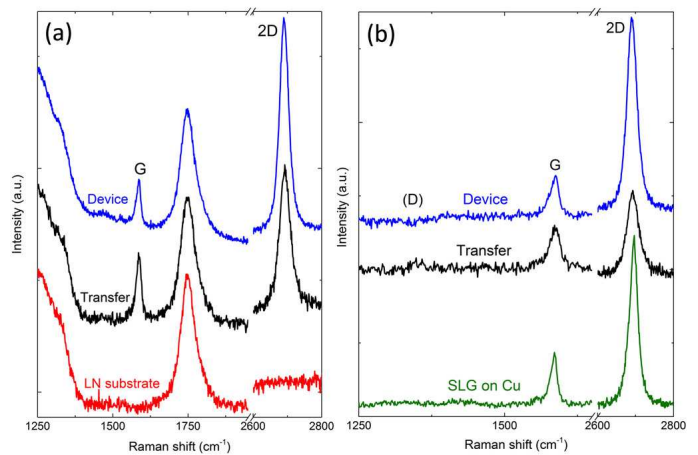}
\caption{(a) Raman spectrum of the bare LN substrate (red curve), of SLG transferred on LN (black curve) and of SLG on LN after device fabrication (blue curve). (b) Raman spectrum of the as-grown SLG on Cu (green curve), of SLG film transferred on LN (black curve) and of SLG on LN after device fabrication (blue curve) obtained after subtracting the substrate contribution.}
\label{fig:Fig5}
\end{figure}
\section{METHODS}
\subsection{Devices fabrication}
\begin{figure*}
\includegraphics[width=145mm]{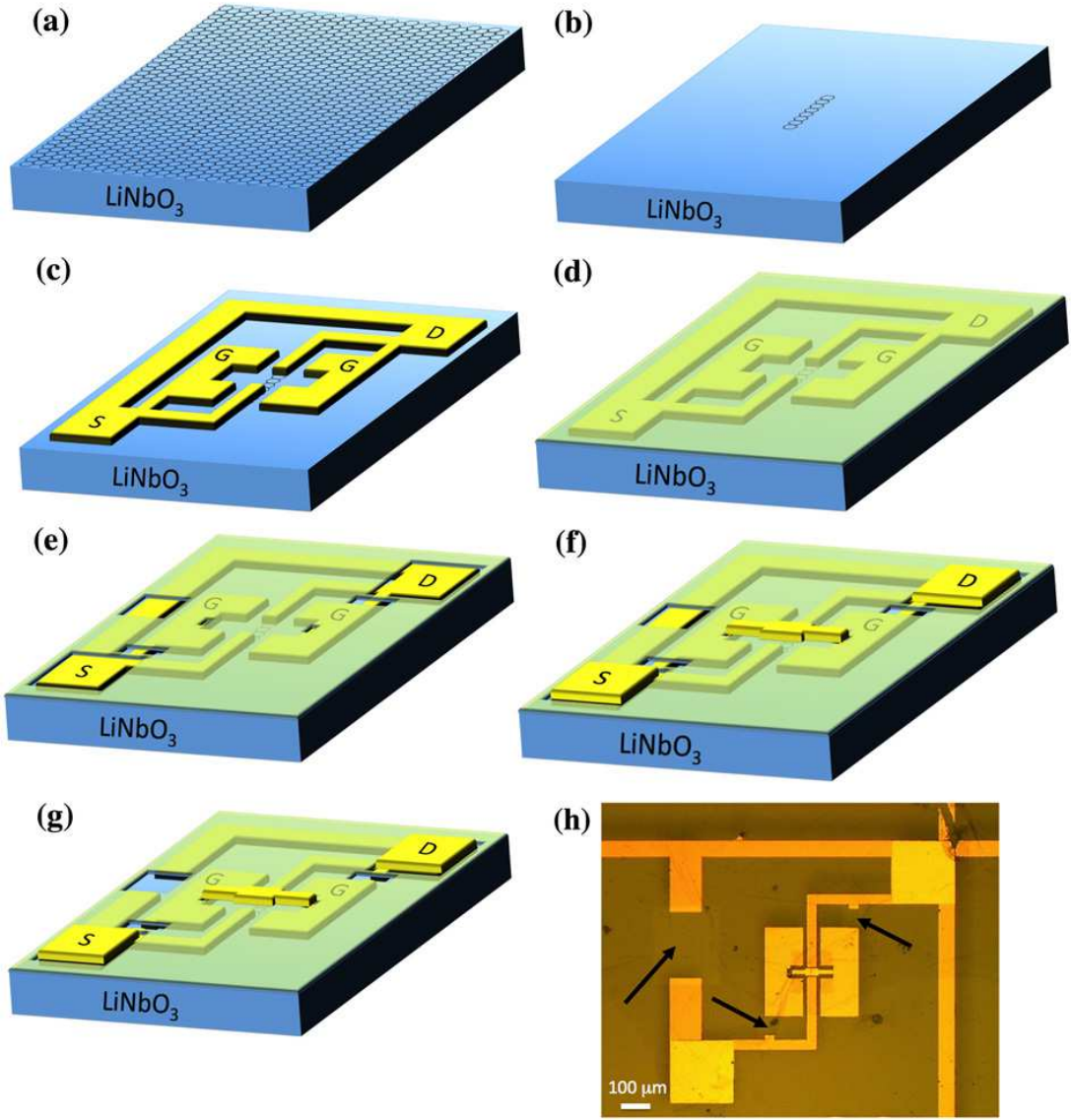}
\caption{(a)-(g) Step-by-step device fabrication process. (h) Optical image after device fabrication. }
\label{fig:Fig6}
\end{figure*}
SLG is grown by chemical vapor deposition (CVD) on 35-$\mu$m-thick Cu following the process described in Ref.\cite{BaeNN2010}. The quality of the material is monitored by Raman spectroscopy using a Renishaw InVia equipped with a 100X objective at 514.5nm, with a laser power below 300$\mu$W to avoid possible heating effects or damage. Fig.\ref{fig:Fig5}b (green curve) shows the Raman spectra of SLG on Cu. The 2D peak is single-Lorentzian, a signature of SLG, with FWHM(2D)=26cm$^{-1}$\cite{FerrPRL2006}. The D to G intensity ratio, I(D)/(G)$\sim$0.1 indicates a defect density$\sim$2.5$\cdot$10$^{10}$cm$^{-2}$\cite{FerrPRL2006, CancNL2011, FerrNN2013, FerrPRB2000}. SLG is then transferred on the positively charged surface of z-cut LN (Roditi International Ltd) by spin coating a 500nm-layer of polymethyl methacrylate (PMMA) and then etching the Cu foil with an aqueous solution of ammonium persulfate\cite{BaeNN2010}. The resulting graphene/PMMA film is rinsed in water and picked up with the target substrate. After drying, the sample is placed in acetone to dissolve the PMMA, leaving a film of SLG on LN. Fig.\ref{fig:Fig5}a plots the Raman spectrum after transfer on LN (black curve). The spectrum for the bare LN substrate is also reported (red curve). The D peak region at$\sim$1350cm$^{-1}$ is convoluted with an intense band at 1200-1450cm$^{-1}$ arising from optical phonons in LN\cite{BarkPR1967}. An additional LN peak is also present at 1744 cm$^{-1}$, which does not overlap with any of the characteristic features of SLG\cite{BarkPR1967}. In Fig.\ref{fig:Fig5}b we plot the transferred SLG spectrum (black curve) after subtracting the reference contribution of the LN substrate, and normalization to the LN peak at 1744cm$^{-1}$. The 2D peak is single-Lorentzian with FWHM(2D)$\sim$38cm$^{-1}$. The position of the G peak, Pos(G), is 1588cm$^{-1}$, while its full width at half maximum, FWHM(G), is$\sim$14cm$^{-1}$. The position of the 2D peak, Pos(2D), is 2691cm$^{-1}$. The 2D/G peak intensity and area ratios, I(2D)/I(G) and A(2D)/A(G), are 2.1 and 4.1, respectively, suggesting a doping concentration$\sim$5$\cdot$10$^{12}$cm$^{-2}$ ($\sim$300meV)\cite{DasNN2008, BrunACSN2014, BaskPRB2009}. The spectra show I(D)/(G)$\sim$0.22, corresponding to a defect concentration$\sim$5.4$\cdot$10$^{10}$cm$^{-2}$\cite{FerrPRL2006, CancNL2011, FerrNN2013, FerrPRB2000}, similar to that before transfer, indicating that negligible extra defects are introduced during the transfer process\cite{FerrPRL2006, CancNL2011, FerrNN2013, FerrPRB2000}.

The fabrication of top-gated GFETs on LN presents additional challenges compared to Si/SiO$_2$. Owing to the pyroelectric nature of LN, a significant static charge can build on both surfaces. To preserve our devices from discharge-induced damage, we initially prepare all metallic features on the LN surface as an electrically connected pattern, i.e. source, drain and floating gate contacts of the GFET are shorted together by means of metallic lines. In this configuration the device can undergo all the required high-T (up to 120$^\circ$C) processing steps without failing. The shorts are then removed in the last step, when no further heating is required aside from normal sensor operation. The device fabrication process is outlined in Fig.\ref{fig:Fig6}. First, SLG channels are patterned (Fig.\ref{fig:Fig6}b) using optical lithography and dry etching in O$_2$ (20W for 20s). A second lithographic step defines the metal contacts (source and drain), as well as the floating gate pads directly in contact with the substrate (Fig.\ref{fig:Fig6}c). Note that all features are shorted together as explained above. Before the deposition of a 40nm-thick Au layer via thermal evaporation, a mild Ar plasma (0.5W, 20s) is used on the exposed SLG areas. This is crucial to achieve a good contact resistance ($<$100$\Omega$), as defects induced by the plasma ensure a good bonding with the metal\cite{ChoiJAP2011}. Further, a 10nm Al$_2$O$_3$ layer is deposited by atomic layer deposition (ALD) at 120$^\circ$C, to serve as gate dielectric (Fig.\ref{fig:Fig6}d). 2nm Al is used as seed layer for ALD\cite{KimAPL2009}. Optical lithography is again used to define apertures in the Al$_2$O$_3$ in order to expose the contact pads (source and drain), part of the shorting lines and a small section of the lateral pads where the top electrode needs to be anchored. The Al$_2$O$_3$ is then wet-etched in an alkaline solution (D90/H$_2$O 1:3) for$\sim$6 minutes, leaving the structure in Fig.\ref{fig:Fig6}e. Another lithographic step is then used to finalize the top-gate structure via thermal evaporation and lift-off of 2/60 nm of Cr/Au. Bonding pads are also prepared in this step, overlapping with those deposited with the contacts (Fig.\ref{fig:Fig6}f). Finally, the electrical shorts are removed with a last lithographic step followed by wet-etching of the Au lines in an aqueous solution of KI/I$_2$ (Fig.\ref{fig:Fig6}g). An optical picture of the final device is shown in Fig.\ref{fig:Fig6}h, the arrows indicating where the Au shorts have been etched. Fig.\ref{fig:Fig5}a shows the Raman spectrum after the device fabrication (blue curve). The laser spot is in the gap between a contact and the top-gate. LN peaks (1200-1450cm$^{-1}$ and 1744cm$^{-1}$) appear as a signature of the underneath substrate\cite{BarkPR1967}. The spectrum after subtracting the LN reference is presented in Fig.\ref{fig:Fig5}b (blue curve) and compared with graphene after transfer (i.e. before fabrication, black curve). The signature of SLG is again confirmed from the single-Lorentzian line-shape of the 2D peak with FWHM(2D)$\sim$31cm$^{-1}$\cite{FerrPRL2006}. Pos(G)=1584cm$^{-1}$, FWHM(G)$\sim$16cm$^{-1}$ and Pos(2D)=2689cm$^{-1}$. I(2D)/I(G) and A(2D)/A(G) are 4.2 and 8.5, respectively, indicating a doping level$<$200meV ($<$10$^{12}$cm$^{-2}$)\cite{DasNN2008, BrunACSN2014, BaskPRB2009}. The doping due to Al$_2$O$_3$ encapsulation\cite{HoACSN2012} is consistent with the value extracted from the electrical data in Fig.\ref{fig:Fig3}. From I(D)/I(G)$\sim$0.16, a defect concentration$\sim$4$\cdot$10$^{10}$ cm$^{-2}$ is obtained\cite{FerrPRL2006, CancNL2011, FerrNN2013, FerrPRB2000}. This indicates that the fabrication process does not introduce defects\cite{FerrPRL2006, CancNL2011, FerrNN2013, FerrPRB2000}.
\subsection{Device Characterization}
T-dependent electrical characterization is performed with a Cascade probe station with a T-controlled chuck, coupled to an HP4142B source-meter. The spectral density of the current fluctuations (S$_I$) is the Fourier transform of the drain current recorded during 100s with a sampling of 1ms. The normalized S$_I$/I$^2$ has the same 1/f dependence for all drain voltages.

The devices are then illuminated by a linearly polarized quantum cascade laser with a frequency range from 1000 to 1610cm$^{-1}$ ($\sim$6.2-10$\mu$m) scanned using a motorized xyz-stage. The laser is modulated using a chopper and the current measured with a current pre-amplifier and lock-in amplifier. The light polarization is controlled with a ZnSe wire grid polarizer. The light is focused using ZnSe lenses with NA$\sim$0.5. The power for each frequency is measured using a bolometric power meter and the photocurrent spectra are normalized by this power to calculate the responsivity.
\subsection{Simulations and Models}
Optical calculations are performed with a finite-difference time-domain (FDTD) method\cite{Tafl2005, LidoJAP2007} assuming an infinite array of infinitely long Au fingers (40nm-thick) on a semi-infinite LN substrate. Thermal transient calculations are performed with a finite element method (FEM)\cite{Comsol} assuming a 500$\mu$m-thick LN substrate on a 3mm-thick Au block (heat-sink) whose back surface is kept fixed at room temperature.

Fig.\ref{fig:Fig2}e shows that, when a chopper induces a periodicity in the illumination (and associated heating), the amplitude of the resulting periodic photoresponse scales linearly with the inverse of its frequency. We present here a thermal model and associated simulations to explicitly study the time-dependent heating and cooling properties of our device and confirm the above conclusion.
\begin{figure}
\includegraphics[width=80mm]{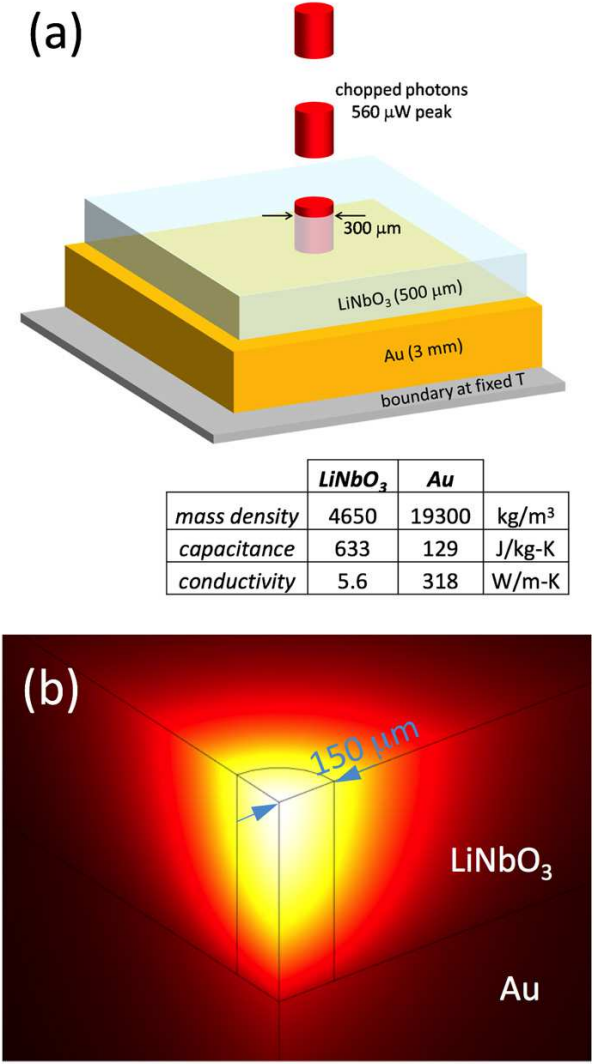}
\caption{(a) Model used for the simulations and associated parameters. (b) Simulated temperature distribution around A 150$\mu$m-radius laser spot (white=hotter, red=colder).}
\label{fig:s1}
\end{figure}
\begin{figure*}
\includegraphics[width=175mm]{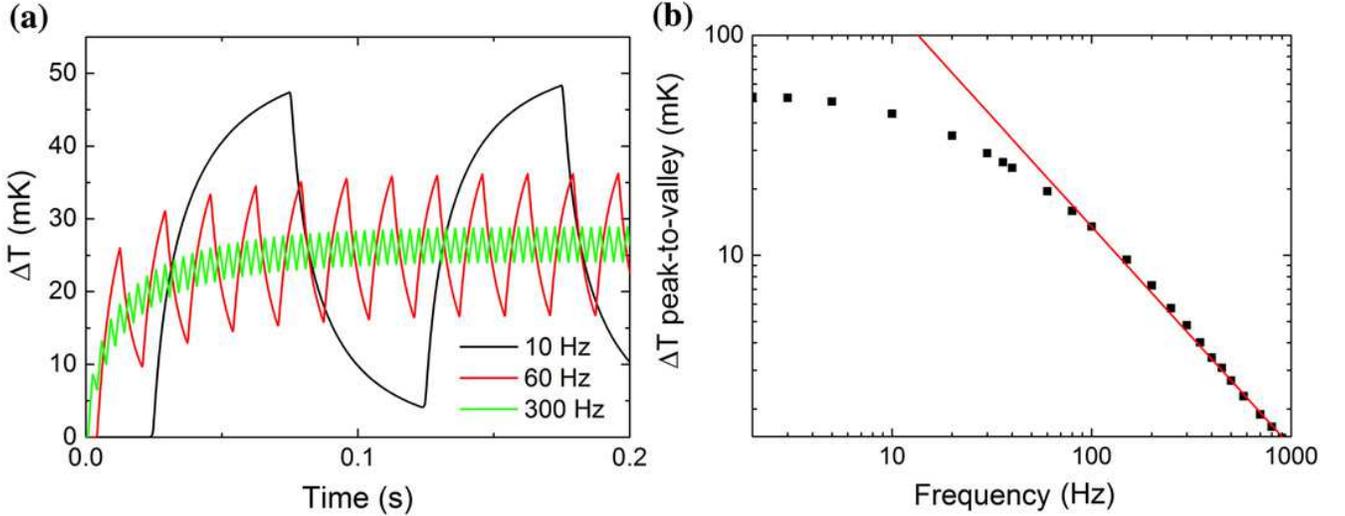}
\caption{(a) Total T transient for 3 chopper frequencies. (b) $\Delta$T peak-to-valley dependence as a function of the chopper speed.}
\label{fig:s2}
\end{figure*}

Fig.\ref{fig:s1}a illustrates how we model the experiment in Fig. \ref{fig:Fig2}e. A laser beam with equivalent characteristics (560$\mu$W, 300$\mu$m in diameter) is chopped at variable frequencies with a 50$\%$ duty cycle. The 500$\mu$m-thick LN substrate is in contact with a 3mm-thick Au layer (that represents the chip carrier), acting as heat sink towards a boundary at fixed temperature T$_0$. Typical materials parameters are reported in Fig.\ref{fig:s1}a\cite{Roditi}. We make the following approximations: 1) Real devices with patterned Au pads reflect$\sim$50$\%$ of the impinging light. We thus assume 50$\%$ reflectance for the LN film and 0$\%$ transmission (i.e. 50$\%$ absorption); 2) We assume radiation uniformly absorbed in the LN film; 3) we disregard heating losses by convection and radiation. Under these assumptions, the T distribution near the laser spot for an arbitrary instant in time is shown in Fig.\ref{fig:s1}b.

Fig.\ref{fig:s2}a shows the simulated temperature transient $\Delta$T=T-T$_0$ for the point at the top of the LN film and at the centre of the illumination spot. The average T saturates at$\sim$T$_0$+50mK for all frequencies, because this would be the equilibrium value for constant illumination at 280$\mu$W (the impinging power is 560$\mu$W/2 at all frequencies). The photoresponse measured by the lock-in in Fig.\ref{fig:Fig2}e overlooks this average $\Delta$T and reflects the peak-to-valley $\Delta$T of a single period, in phase with the chopper. This, on the other hand, varies significantly in Fig.\ref{fig:s2}a for the three frequencies reported. The full dependence of the $\Delta$T peak-to-valley on f is shown in Fig.\ref{fig:s2}b. The calculations are performed with a finite element method (FEM)\cite{Comsol} using the assumptions made for Fig.\ref{fig:s1}. Despite the approximations, our simulations reproduce the 1/f behavior observed experimentally above 60Hz in Fig.\ref{fig:Fig2}e. For low frequencies ($<$10Hz), the system saturates within one period.

\begin{figure}
\includegraphics[width=80mm]{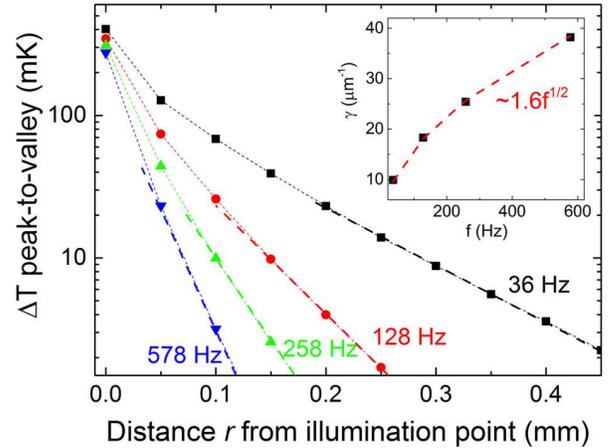}
\caption{Radial decay of $\Delta$T peak-to-valley for different f.}
\label{fig:s3}
\end{figure}

To match the conditions used for the photomapping in Fig.\ref{fig:Fig2}a-d, we now reduce the spot diameter to 10$\mu$m and increase the laser power to 1.8mW. Fig.\ref{fig:s3} shows the $\Delta$T peak-to-valley for a number of points on the LN surface as a function of their distance, r, from the centre of the illumination spot. Outside the spot, $\Delta$T decays with a rate that depends on f. A decay of the form$\sim$exp(-$\gamma$r) is found, with $\gamma$=1.6f$^{1/2}$. This means that the higher f, the less the heat delivered by the laser in one cycle is allowed to spread radially. For this reason, the photomapping in Fig.\ref{fig:Fig2}d is much more resolved than in Fig.\ref{fig:Fig2}a.
\begin{figure*}
\includegraphics[width=160mm]{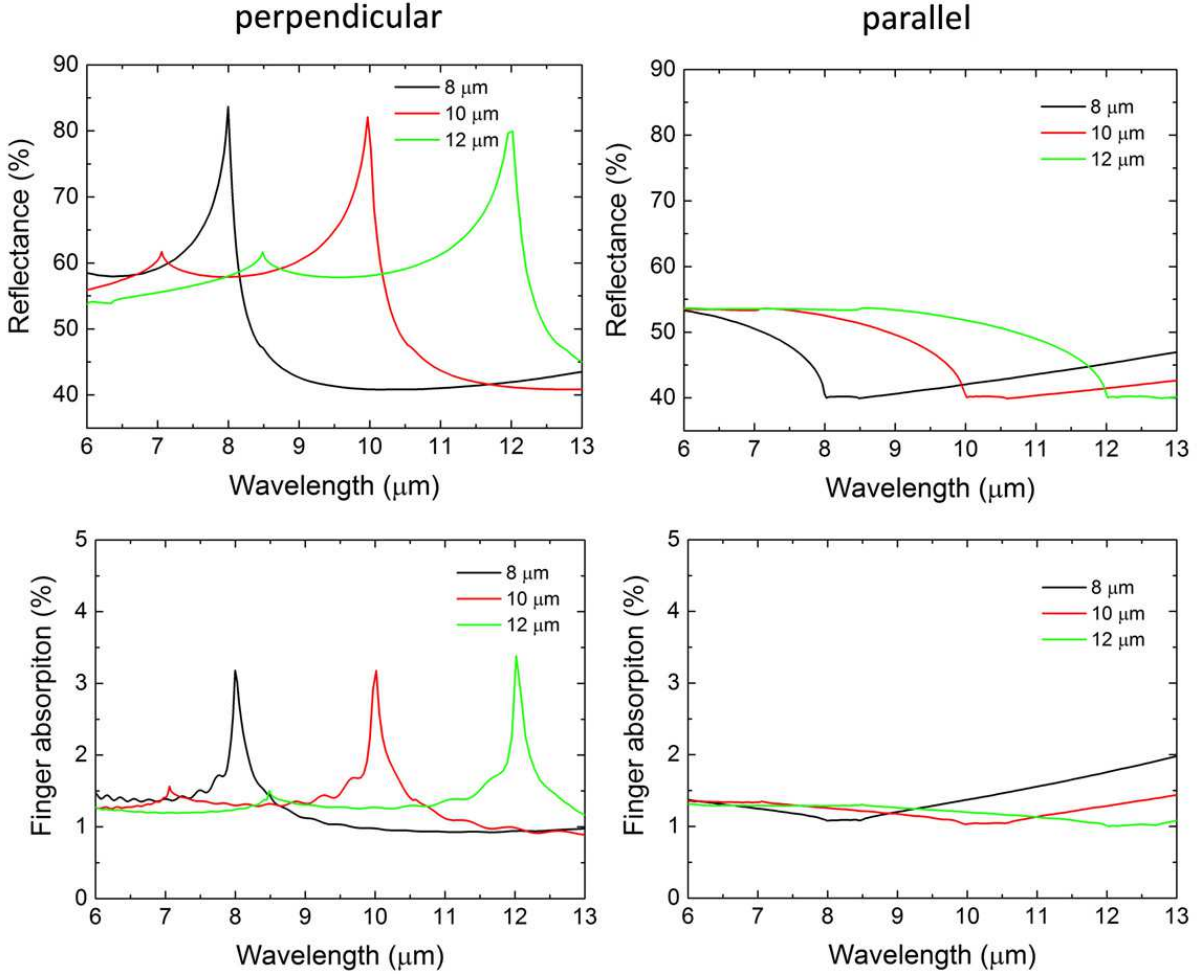}
\caption{Simulated reflectance and absorption in the Au fingers for parallel and perpendicular polarized light, for different finger pitch.}
\label{fig:s4}
\end{figure*}
\subsection{Wavelength-selective absorption}
Fig.\ref{fig:Fig2}f indicates that the photoresponse of a device with lateral pads patterned as parallel fingers with pitch L and filling ratio 0.5 exhibits a peak in the ratio between parallel and perpendicular incident light. The spectral position of such peak is linked to the geometrical parameter L. Fig.\ref{fig:Fig2}g shows that the total absorption (linked to the device photoresponse) has peaks in the parallel/perpendicular ratio at wavelengths that closely match the experiment. However, Fig.\ref{fig:Fig2}g does not allow to discriminate between reflectance, pattern absorption and substrate absorption, i.e. the individual components contributing to the total absorption.

Fig.\ref{fig:s4} plots the simulated reflectance and absorption in the Au fingers for both parallel and perpendicular polarized light for L=8, 10, 12$\mu$m, as in Fig.\ref{fig:Fig2}f,g. For both reflectance and finger absorption, peaks are found at the corresponding wavelength. Reflectance and finger absorption are measured as $\%$ of the total incident light. This means that, even if more light is reflected at resonance for perpendicular polarization, this is also the condition that results in maximum absorption in the fingers. However, peak absorption in the fingers at resonance is$\sim$3$\%$ for perpendicular polarization (versus$\sim$1$\%$ off-resonance and for parallel polarization), while the measured absorption of a 500$\mu$m-thick LN substrate is$\sim$75$\%$ in this wavelength range. Hence, substrate absorption dominates, overshadowing the absorption features in Fig.\ref{fig:s4}, and the total absorption is determined by 1-RFL. In the limit of thin substrates, however, which is the case of highest technological interest, the photoresponse will strongly depend on the absorption in the fingers, and simulations like that in Fig.\ref{fig:s4} will be crucial to optimize the design of patterns to enhance absorption and spectral selectivity.
\subsection{Characterization of the direct effect}
We now investigate the conductivity modulation induced by the substrate (direct effect) in devices where the floating top-gate is omitted. We deposit the Al$_2$O$_3$ dielectric layer in all cases, to improve device stability. We extract the TCR by means of thermo-electrical measurements, as in Fig.\ref{fig:Fig3}b, but with T ramp expanded to 1$^\circ$C, because of the weaker response.
\begin{figure}
\includegraphics[width=80mm]{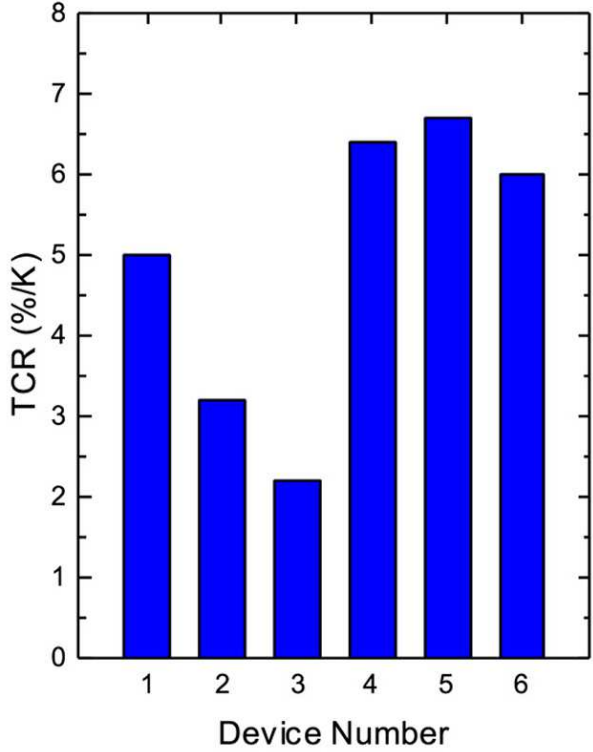}
\caption{Measured TCR for devices on LN exploiting the direct effect.}
\label{fig:s5}
\end{figure}

Fig.\ref{fig:s5} plots the TCR distribution across multiple devices with equivalent geometry. The average TCR (in module) is 4.9$\%$/K, on par with that reported by Ref.\cite{HsieAPL2012} for SLG on lead titanate zirconate ($\sim$6$\%$/K). Thus, the contribution of the direct effect to the device sensitivity is negligible compared to the amplified response achieved with the floating gate, which remains dominant even for the smallest pixel size ($A_{C3}/A_{C2}$=1, see Fig.\ref{fig:Fig4}a). This means that, in most practical cases, there is little benefit in placing SLG in direct contact with the pyroelectric substrate. We envisage that the performance of our devices can be further optimized by implementing a suitable spacer between SLG and the pyroelectric substrate, chosen from materials known to provide a smooth and inert interface with SLG (e.g. boron nitride\cite{DeanNN2010}).
\subsection{Device resilience to large temperature variations}
The breakdown field of the Al$_2$O$_3$ dielectric layer limits the dynamic range of the GFET, for a given TCR. We consider safe to apply gate fields up to 5 MV/cm, corresponding to an applied voltage of $\pm$5V for a 10-nm-thick alumina film. Close inspection of Figs.\ref{fig:Fig3}a,b reveals that, if 0.2$^\circ$C induce a gate voltage of -0.44V, a T variation$>$3$^\circ$C may result in device failure. All samples are subjected to much larger ($>$10$^\circ$C) T variations. Surprisingly, the devices are very resilient, with no accidental failure and consistent response.
\begin{figure}
\includegraphics[width=80mm]{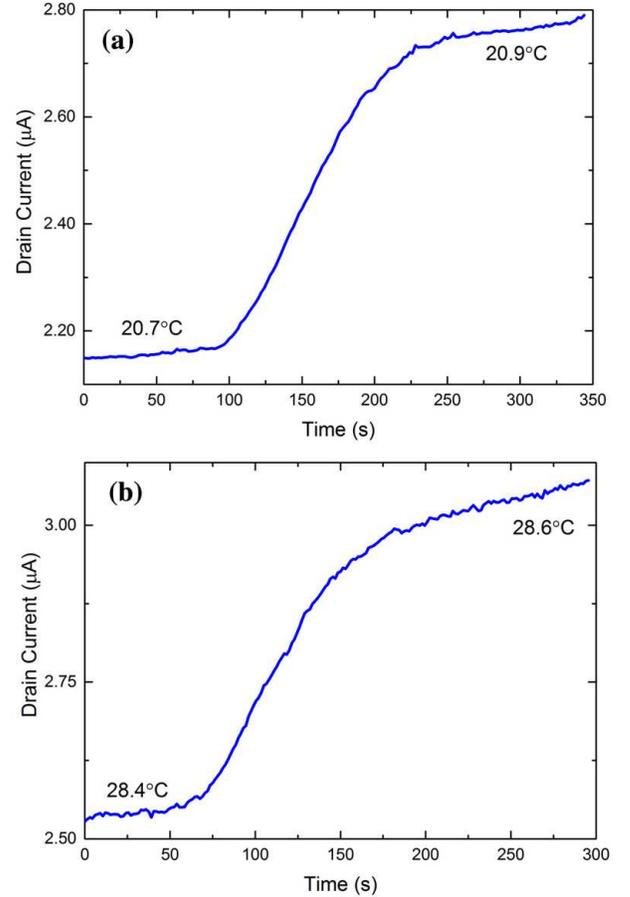}
\caption{Drain current response for (a) T increase of 0.2$^\circ$C from RT and (b) after a slow 8$^\circ$C heating ramp at 2$^\circ$C/hour.}
\label{fig:s6}
\end{figure}

The reason for this behavior resides in the T gradient, i.e. how slow or fast the device experiences the T variation. Fig.\ref{fig:s6}a plots the response of a representative device with TCR$\sim$130$\%$/K as T increases by 0.2$^\circ$C in 3 minutes. This corresponds to an induced gate voltage of -0.22V (because of the different $A_{C3}/A_{C2}$ ratio, the TCR is roughly half that in Fig.\ref{fig:Fig3}b). Accordingly, increasing T by 8$^\circ$C should result in an induced voltage$\sim$9V, destroying the device. The T is thus slowly raised by$\sim$8$^\circ$C with (2$^\circ$C/hour) and the response to a T variation of 0.2$^\circ$C is measured again (Fig.\ref{fig:s6}b). Interestingly, not only the device survives the dielectric breakdown, but its response is still consistent with Fig.\ref{fig:s6}a. Indeed, the drain current at 28.4$^\circ$C (2.53$\mu$A, Fig.\ref{fig:s6}b) is almost the same as previously measured at$\sim$20.8$^\circ$C (Fig.\ref{fig:s6}a), indicating that the  operating point tends to drift in the opposite direction of the applied stimulus. This is not due to the GFET hysteresis, as this would have been evident also in Fig.\ref{fig:Fig3}b (where no hysteresis is observed). Hence, we conclude that the drift preserving the device comes from the slow ($\sim$hours) internal discharge of the pyroelectric crystal. This is negligible over a few minutes (Fig.\ref{fig:Fig3}b), but can play an important role over several hours.
\subsection{Drain current response for large pixels}
\begin{figure}
\includegraphics[width=85mm]{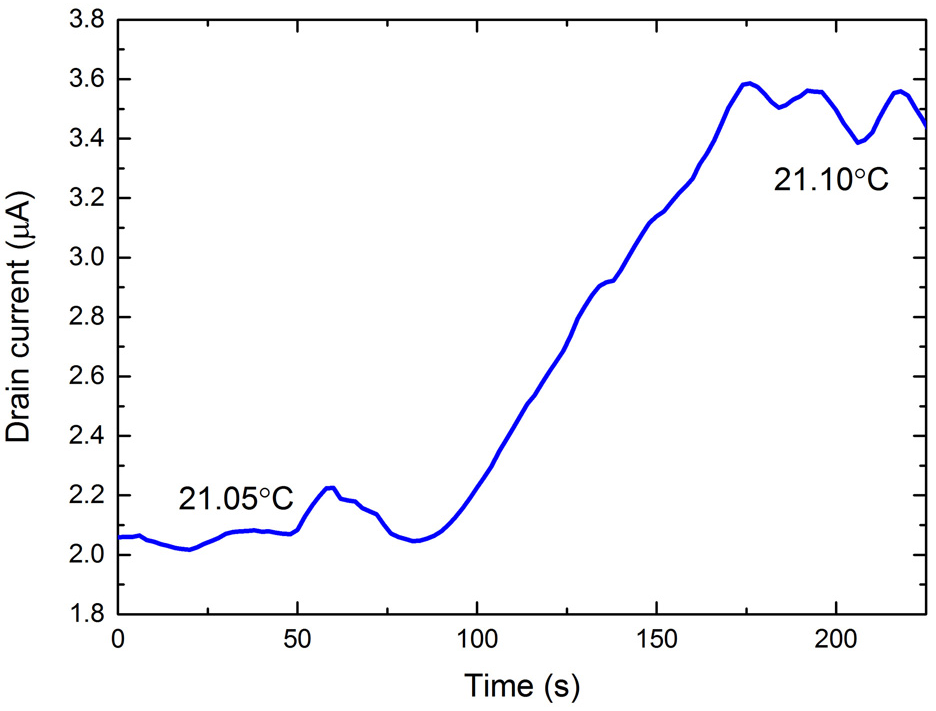}
\caption{Drain current response for a 300x300$\mu$m$^2$ pixel upon a 0.05$^\circ$C T increase.}
\label{fig:s7}
\end{figure}
Here we report the TCR measured on a 300x300$\mu$m$^2$ pixel. By applying a drain voltage$\sim$10mV, the drain current is measured while changing T, as for Fig.\ref{fig:Fig3}b. However, due to the large TCR, we increase T$\sim$0.05$^\circ$C (rather than 0.2$^\circ$C, as for Fig.\ref{fig:Fig3}b). Fig.\ref{fig:s7} reports the drain current response, showing a TCR$\sim$900$\%$/K.
\subsection{Acknowledgements}
We acknowledge funding from EU Graphene Flagship, ERC Grant Hetero2D, and EPSRC Grant Nos. EP/ 509 K01711X/1, EP/K017144/1, EP/N010345/1, EP/M507799/ 5101, and EP/L016087/1. We thank Graphenea for the provision of some CVD graphene samples on LN.

\end{document}